\documentclass{llncs}
\usepackage{makeidx}  % allows for indexgeneration
\usepackage{graphicx}
\usepackage{slashbox}
\usepackage{cite}
\usepackage{caption}
\usepackage{stmaryrd}
\begin{document}

\frontmatter          % for the preliminaries
\title{Fingerprinting Cryptographic Protocols with Key Exchange using an Entropy Measure}
%
%                                     also used for the TOC unless
%                                     \toctitle is used
%
\author{Shoufu Luo\inst{1} \and Sven Dietrich\inst{1,2}}
\authorrunning{Luo et al.} % abbreviated author list (for running head)
%
%%%% list of authors for the TOC (use if author list has to be modified)

\institute{The Graduate Center, City University of New York, \\
\email{sluo2@gradcenter.cuny.edu}
  \\
\and John Jay College of Criminal Justice, City University of New York, 
\email{spock@ieee.org}
}

\maketitle              % typeset the title of the contribution

\begin{abstract}
Encryption has increasingly been used in all applications for various purposes, but it also brings big challenges to network security.  In this paper, we take first steps towards addressing some of these challenges by introducing a novel system to identify key exchange protocols, which are usually required if encryption keys are not pre-shared. We observed that key exchange protocols yield certain patterns of high-entropy data blocks, e.g. as found in key material. We propose a multi-resolution approach of accurately detecting high-entropy data blocks and a method of generating scalable fingerprints for cryptographic protocols. We provide experimental evidence that our approach has great potential for identifying cryptographic protocols by their unique key exchanges, and furthermore for detecting malware traffic that includes customized key exchange protocols. 
\end{abstract}

\section{Introduction}

%The arms race between malware creators and security researchers has been ongoing. While security researchers develop various detection techniques, new ways have been found by malware creators to defeat those mechanisms.  For example, HTTP has been widely adopted by botnets as a primary transport vehicle to hide malicious activity in the huge volume of innocent traffic and to fool enterprise firewalls. To avoid single-point failure of a centralized C\&C server, peer-to-peer topologies boost the resilience of the command and control (C\&C) infrastructure, and also conceal the C\&C traffic among numerous popular peer-to-peer applications, e.g. filesharing. However, t

In the network security field, the use of encryption for malicious purposes brings new challenges to network security defense. For example, encryption has prevented botnet traffic from being inspected and detected by defense systems based on deep-packet inspection (DPI), which used to be very effective up to that point. For symmetric encryption and decryption, a secret key $k$ shared among two communicating parties is required, either pre-shared or negotiated on the fly using cryptographic key-exchange protocols. Most common cryptographic protocols \cite{ssl:rfc6101, tls:rfc5246, ssh:rfc4251} using symmetric encryption to secure the channel use a key exchange protocol, such as the Diffie-Hellman key exchange \cite{dhe:rfc2631}. %\cite{ssl:rfc6101, tls:rfc5246}, SSH %\cite{ssh:rfc4251} etc.

\begin{figure}[h]
	\centering
	\includegraphics[scale=0.8]{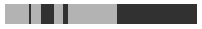}
	\caption{Visualization of Entropy Distribution: dark portions are high-entropy blocks.}
	\label{fig:demo}
\end{figure}

Depending on the protocol design, key material is distributed differently along the traffic stream. As key material has high entropy compared to normal traffic, the traffic for the key exchange exhibits detectable characteristics, namely the uniqueness of the distribution of key material allowing for proper discriminating characteristics, as shown in Figure \ref{fig:demo}.
 %With botnets exhibiting such behavior, this significantly increases the chance of detecting these characteristics. 
Using an entropy metric, it may not be hard to test the hypothesis whether a byte string is ``random,'' if that byte string is sufficiently long. The problem becomes harder if the given string is relatively short, i.e. undersampled, or if the goal is to identify which part of the string contains random bytes, in particular, deciding the boundaries of those random bytes (also known as blocks of interest). It is therefore challenging to characterize a stream by the distribution of embedded random bytes, or so-called high-entropy blocks. 

To avoid being treated as an anomaly, malware might try to use standard cryptographic protocols (e.g. SSL/TLS) for secure communication, effectively preventing DPI. However, standard protocols such as SSL can potentially be subject to a man-in-the-middle attack. However, malware in general tends to avoid using standard protocols and instead employs a customized variant. Only 10\% of malware utilize TLS as a form of encryption, according to a recent study \cite{anderson2016deciphering}. To ensure fresh key material, a new key exchange is desirable for every new command-and-control (C\&C) session of the malware \cite{dd2008:malware, art:stormnugache}.

Our work offers a systematic way to characterize network traffic through key exchange behaviors and generate scalable fingerprints based on detected high-entropy blocks. The system mainly consists of two parts: the high-entropy block detection and the fingerprint generation. First, we aim to identify high-entropy blocks from a traffic stream using sample entropy via a sliding window. Second, with all high-entropy blocks identified, entropy-based fingerprints for network flows will be generated by the distribution of high-entropy blocks. Our contribution also includes:
\begin{itemize}
\item A new method of identifying cryptographic protocols, raising the bar for malicious activities that abuse customizing cryptographic protocols to evade inspection.
\item A voting mechanism that efficiently boosts the accuracy of entropy estimation when undersampled using a multi-resolution analysis. 
\item A statistical approach to estimate the range of high-entropy data blocks and build scalable entropy-based fingerprints for key exchange protocols in the form of regular expressions.
\end{itemize}
To the best of our knowledge, our work is the first attempt to fingerprint key exchange protocols by the distribution of key material and apply such a technique to malware detection. By design our approach can be implemented and deployed as a standalone system. However, it is not the intention to replace any existing detection techniques, but rather to complement them. This system can be built with existing systems as a plug-in component, in particular those relying on a certain degree of payload analysis, e.g. \cite{tamd}. Moreover, a component of our system can be a useful tool for the security community, e.g. for identifying high-entropy portions of a given data block, such as detection of packed malware binaries.  

\subsubsection{Related Work}

%There are works that use entropy of network traffic for intrusion detection. 
Olivain et al.~\cite{netentropy} proposed to use cumulative entropy of network flows for detection of specific attacking behaviors targeted at known cryptographic protocols, i.e. SSL. Instead of an aggregation, our work aims to fingerprint the entropy distribution along the examined traffic. Our approach is still applicable for their purpose in a more precise way. Meanwhile, we adopt the technique they propose, \emph{N-truncated entropy}, for entropy estimation, which is also used by Dorfinger et al.~\cite{Dorfinger2011TMA} for classifying encrypted and unencrypted traffic.
% measure the entropy of each block (within a chosen window) and identify high-entropy blocks (i.e. key material) as well as low-entropy blocks (i.e. control information).
% We adopt a technique, , developed in \cite{netentropy}, to statistically predict whether a $N$-character string is high-entropy or not. 
%Dorfinger et al.~\cite{Dorfinger2011TMA} adopted such techniques to classify encrypted and unencrypted traffic as well. 
There is prior work \cite{white2013clear} that shows how entropy tests can be used to detect encrypted or compressed packets from network streams. Again, we provide a more reliable mechanism to detect high-entropy areas as one of our essential contributions.  

Our work shares an interest from the field of protocol identification. Most of the work in that field is mainly learning-based, relying on network-observable features \cite{Kara:2005, Wright:2006}. For example, Wright et al. \cite{Wright:2006} proposed to identify the cryptographic protocol of individual encrypted TCP connections using post-encryption observable features, such as timing, size, direction etc. To some extent, our approach can also be also applied for this purpose. However, there are known obfuscation techniques which could be used to evade this, such as obfsproxy\cite{obfsproxy} and FTE\cite{dyer2013protocol}. As discussed in \cite{wang2015seeing}, obfuscation can be detected with entropy-based tests over the packet payloads. Our approach does the same by extracting entropy-based fingerprints.

Zhang et al.\cite{ZhangPM13} proposed to detect encrypted traffic by looking for \emph{N} sequential high-entropy packets of all first \emph{M} packets of one network flow adopting the cumulative entropy technique. In 2015, Zhang et al. \cite{zhang15} improved their previous work by detecting of high-entropy flows as an additional measure to score a host being a bot for BotHunter~\cite{bothunter2007usenix}. Applicable to the same problem, our approach is different from theirs by fingerprinting malware with customized cryptographic protocols, such as Nugache, as will be shown. Unlike their work, our work does not rely on another system for detection.

The rest of this paper is organized as follows. We begin with background on entropy and its estimators. In section \ref{sec:method}, we discuss our methodology in detail, including how to identify high-entropy blocks, a voting mechanism as well as a filtering method for false positives reduction, etc. Following that,  section \ref{sec:eval} presents evaluation and analysis of our approach with three different dataset. Finally, we conclude this study by discussing limitations and directions of future work. 

\section{Background}

\subsection{Entropy}
Introduced by Shannon \cite{shannon1948}, entropy is used as a measurement of the amount of information that is missing before reception. In the context of cryptography, it is used as a measure of randomness (or uncertainty), equating higher entropy with higher randomness. Let $X$ be a discrete random variable under an arbitrary distribution $\mathcal{P}$ on a countable alphabet $\Sigma=\{x_1, ..., x_m\}$. The definition of Shannon entropy can be generally expressed by the equation (\ref{entropy}),
\begin{equation}
\label{entropy}
H(X) = -\sum^m\limits_{i=1} p(x_i) \log_2 p(x_i) % H(P) v.s. H(X), need to clarify 
% H(P) v.s. H(X), need to clarify 
\end{equation}

%For example, a random string uniformly drawn from $\Sigma = \mathrm{\{0h00, ... , 0hff\}}$ would have entropy of 8 for $\forall i$ $p(x_i)=\frac{1}{256}$. It could be also interpreted to 8 bits of entropy, which means all bits of each character are completely random and unpredictable. 

% where $p(x_i)$ is the probability of $x_i$ occurring. 
The entropy $H(X)$ yields a maximum value when all $p(x _i)$ are equal to $\frac{1}{m}$, i.e. uniformly distributed. %, the most uncertain situation. 
In cryptography, as a fundamental requirement of security, key material should have high entropy in order to be hard to predict. % a string generated by such a random variable is considered random and can be used as key materials. %The logarithm base of 2 is chosen for convenience. Choosing other bases will not affect the meaning of this definition. % That's because log_b a = ln a / ln b ;-)

\subsection{Entropy Estimator}
Entropy can be easily obtained by the equation (\ref{entropy}) if given a random variable whose probability distribution is known. However, in practice, $\mathcal{P}$ may remain unknown for most scenarios.
%It leads to the open problem how to estimate the probability of a random variable from its samples. 
%A well-known estimator is the \emph{Maximum Likelihood Estimator (MLE)}.
Frequently, $p(x_i)$ could be still estimated by the relative frequencies of the outcome $x_i$ from a large number of trials. The probability of $x_i$ is thereby $\hat{p}(x_i) = \frac{n_i}{N}$, where $n_i$ is the number of times $x_i$ occurs and $N$ is the total number of trials or samples. Hereby, the \textit{sample entropy}, a.k.a. maximum likelihood estimator (MLE) \cite{Antos2001}, can be estimated as in the equation below.

\begin{equation}
\label{sample}
\hat{H}_N^{MLE}(X) \equiv -\sum^m\limits_{i=1} \hat{p}(x_i) \log_2 \hat{p}(x_i)
\end{equation} 

Even though MLE is an unbiased estimator of $H(X)$ when $N$ tends to infinity where $\hat{p}(x_i)$ approximates $p(x_i)$ and $\hat{H}_N^{MLE}(X)$ approximates real $H(X)$. When N is not sufficiently large, namely \emph{undersampled}, $\hat{H}_N^{MLE}(X)$ highly bias, in particular, $N < m$ or $N \sim m$.  There is no universal rate at which the error of MLE compared to $H(X)$ would be close to zero \cite{Antos2001}. There are attempts that aim to subtract the bias directly, such as the Miller-Madow corrector~\cite{miller}, the Jackknife corrector~\cite{jackknife} and the Paninski corrector~\cite{Paninski03}. However, the bias is still significantly high when  $N < m$ or $N \sim m$. Moreover, it has been proven difficult to find an unbiased estimator~\cite{Paninski03, Schuermann04biasanalysis}. Unfortunately, the Paninski corrector is unbiased but if and only if $\mathcal{P}$ has a uniform distribution, which can not be guaranteed. Furthermore, according to this study \cite{netentropy}, $\hat{H}_N^{MLE}(X) \sim H(X)$ is valid if and only if $N \gg m$, which typically means $N$ is of the order of roughly at least \emph{10} times as large as $m$. In another word, if $\Sigma_0=\mathrm{\{0x00, ..., 0xff\}}$ (i.e. m=$|\Sigma_0|$=256), it would require around 2,000 samples to possibly obtain a reasonable estimated entropy. That makes it impractical for the purpose of profiling network traffic as key material usually is at most hundreds of bytes (256 bytes = 2048 bits). For example, in a typical TLS handshake, a client random number only contains 28 bytes.

\subsection{N-truncated entropy $H_N(X)$} 
%SF-JUL29
Similar to Olivain et  al.~\cite{netentropy}, an accurate entropy value is not of our main focus, but rather the probability of a string being generated from a uniform distribution. The \emph{N-truncated entropy} $H_N(X)$ proposed by Olivain et al. meets our needs, which is the average of the sample entropy $\hat{H}_N^{MLE}(X)$ over all strings of length of $N$ drawn at random from the distribution $\mathcal{P}$, as defined below. % In particular, when $\mathcal{P}$ is the uniform distribution $\mathcal{U}$, its N-truncated entropy is $H_N(\mathcal{U})$.

%~ proposed in their work ,$H_N(\mathcal{P})$, which is defined as the average of the sample entropy $\hat{H}_{MLE}(\mathcal{P}_N)$ over all strings $s$ of length $N$ drawn at random according to the distribution $\mathcal{P}$:
\begin{equation}
\label{netranc}
\small H_N(X)=\sum\limits_{ \Sigma_i n_i =N } \left[ {N \choose {n_0, ... , n_{m-1}}} \prod\limits_{i=0}^{m-1} p_i^{n_i} \left(- \sum^{m-1}\limits_{i=0}\frac{n_i}{N} \log_2 \frac{n_i}{N}\right) \right]
\end{equation}

By construction, $\hat{H}_N^{MLE}(X)$ is an unbiased estimator of $H_N(X)$ for an arbitrary distribution $\mathcal{P}$. More importantly, $\hat{H}_N^{MLE}(X)$ gives a statistical indication that how close the distribution $\mathcal{P}$ is to being uniform by comparing to $\hat{H}_N^{MLE}(W)$ given that $W$ be a random variable under a uniform distribution $\mathcal{U}$. In section \ref{highent}, we describe how to obtain both values. Alternatively, if a string $s$ of length $N$ with each sample drawn from $\mathcal{P}$, we use $\hat{H}^{MLE}_N(s)$ instead of $\hat{H}_N^{MLE}(X)$. To differentiate this, $w$ is used instead if uniform distribution, $\mathcal{U}$. $H_N(X)$ has an upper bound of $log_2 \min\{m, N\}$ as it reaches its maximum value if all $\hat{p}_{x_i}$ are equal, either $\hat{p}_{x_i} = \frac{1}{N}$ if $N < m$ or $\hat{p}_{x_i} = \frac{1}{m}$ otherwise. In either case, uncertainty reaches its maximum. %$\mathcal{P}$ turns out to be uniform. 

\section{Methodology}
\label{sec:method}
In this section, we discuss in detail the techniques we used and developed, accompanied by experimental evidence. 

\subsection{Sliding Window}
%\lable{slidingwin}
To obtain entropy information of different portions within the traffic stream, a sliding window moves over the traffic with a step of one byte while sample entropy will be measured for each chunk of bytes in that window. Bytes in each window  form a block.

%Noted, assuming $\Sigma_0$ be the alphabet, there N samples in this case will be used for estimation. Figure \ref{fig:entropyd} shows the plot of sample entropy of the first 384 bytes from a randomly captured TCP stream using a 32-byte sliding window. There peaks and valleys in Figure \ref{fig:entropyd} indicate various levels of randomness of each portions in the traffic stream. %Also, given N=32, $\hat{H}_N^{MLE}(s) \leq 5$ as previously discussed.
%\begin{figure}[h]
%	\centering
%	\includegraphics[scale=0.25]{./figure/blkplot.png}
%	\caption{Sample Entropy Output with a 32-byte window }
%	\label{fig:entropyd}
%\end{figure}

%estimate $\hat{H}^{MLE}_N(s)$ for each string $s$ falling inside the window. That way we test whether $s$ is a random string, i.e. a high-entropy block,or not, by comparing to $\hat{H}^{MLE}_N(w)$, where $w$ is drawn from a uniform distribution. %The figure \ref{fig:swin} illustrates this procedure and

% \begin{figure}[h]
% 	\scriptsize
% 	\includegraphics[scale=0.55]{./figure/slidewin.png}
% 	\caption{\scriptsize A Sliding Window over Traffic Stream}
% 	\label{fig:swin}
% \end{figure}

%\end{figure}

\begin{figure}[h]
	\centering
	\includegraphics[scale=0.40]{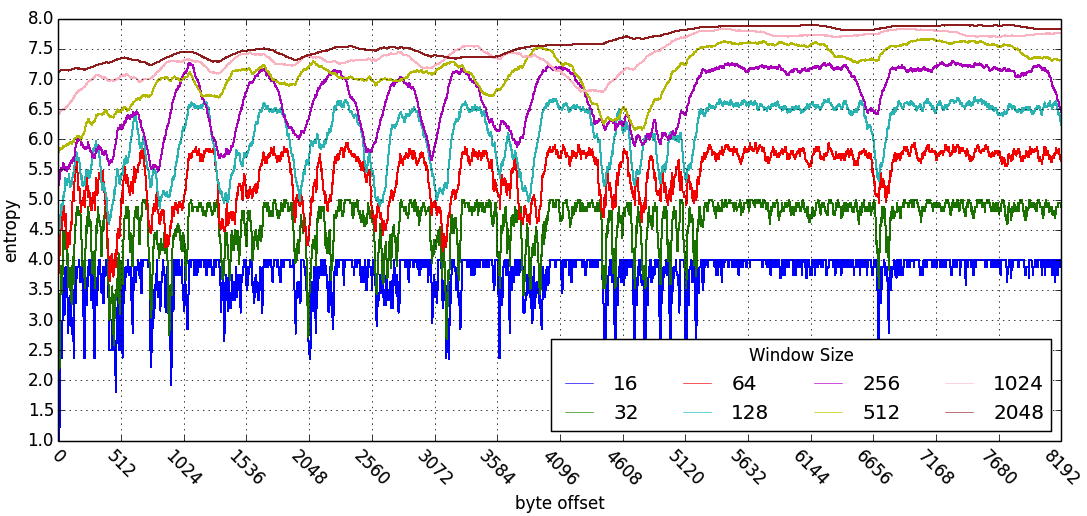}
	\caption{Entropy plot of a TLS sample traffic using different sliding window sizes, from bottom to top (-byte): 16, 32, 64, 128, 256, 512, 1024 to 2048.}
	\label{fig:winsz}
\end{figure}

The window size determines sample size, which directly impacts the accuracy of sample entropy. If the sample size is too small, the sample entropy might not be accurate enough to be meaningful. Equation (\ref{noiseprob}) roughly estimates the probability of a N-byte string appearing to be ``random'', i.e. each char in the alphabet only occurs once in the string. Fix $\Sigma$ to be $\Sigma_0$ and then let $m$=256. Let $N$=16 be a 16-byte sliding window.  Pr[X=e]=0.6197. That is, there is a 40\% probability that an arbitrary string appears random, i.e. a forty percent chance of a false positive. However, if N=32, Pr[X=e]=0.082. This confirms the discussion in Paninski et al. \cite{Paninski03} that one should never use less than 16 bytes for entropy estimation when $\Sigma_0$ is used. 
\begin{equation}
Pr[X=e]=1\cdot \frac{m-1}{m} \cdot ... \cdot \frac{m-N+1}{m} = \prod\limits_{i=0}^{N-1} \frac{m-i}{m} 
\label{noiseprob}
\end{equation}
If the sliding window grows to be too large, it is likely to mix high-entropy areas with low-entropy areas, confusing the difference between them.  
%In another word, the boundary between a high-entropy area and a low-entropy area vanishes. 
As shown in Figure \ref{fig:winsz}, when the window size is small, e.g. 16-byte, the curve is fuzzy and has too many valleys (low-entropy) and peaks (high-entropy), while as the window size goes larger, e.g. 1024 or 2048-byte, the curve becomes flatter and valleys or peaks are not distinctive anymore. 

%Statically speaking, a smaller window is more likely to introduce more false positives, i.e. falsely identified high-entropy blocks, whereas a larger window may fail to detect some high-entropy blocks, i.e. produce false negatives. 
%To be more specific, we can roughly estimate the probability of false positives as follows: let $X$ be the event of N-character strings and $e$ be an event of a non-random string appearing to be ``random'', i.e. high-entropy. 
%We will also use the term `noise' to refer to false positives in the rest of this paper. 

A smaller window is more likely to mistakenly identify a non-random data area to be ``random'' (false positive), while a larger window possibly fails to identify real high-entropy area (false negative). 
%Either case is not ideal for differentiating high-entropy areas from low-entropy ones. 
The choice of window size will heavily depend on the minimum length of key materials of interest. In case of TLS, we choose a 32-byte sliding window as it is good for the minimum length of interests, i.e. 28-byte client random number. % In other scenarios, one may choose accordingly. 
In summary, as the window slides over the data with a one-byte step, each block is labeled as either high-entropy or low-entropy. A list of consecutive either high-entropy blocks or low-entropy blocks then forms a \emph{unit}, more precisely a high-entropy unit or a low-entropy unit respectively. 
%The score will be then normalized to one if the bytes within the sliding window contains high entropy otherwise to zero. %The details of normalization will be discussed in later sections. 

\subsection{Baseline $H_N(\mathcal{U})$}
\label{highent}

%To compare $\hat{H}^{MLE}_N(s)$ to $\hat{H}^{MLE}_N(w)$

To identify a high-entropy block, we follow the idea used by \cite{netentropy}, i.e. the Monte-Carlo method, as it provides a level of confidence of a string being random. We first repeatedly generate strings of length of $N$ with each byte sampled from a random source, e.g. \emph{/dev/urandom} on MacOS X. Then, we calculate the mean $\mu$ and standard deviation $\sigma$ of sample entropy using all samples. Here, $\mu$ and $\sigma$ summarize the distribution of the sample entropy of random strings of length $N$. By a specific number $t$ of standard deviations, we can obtain the proportion of sample strings falling within the range of $\mu \pm  t\times \sigma$. This proportion provides us with a confidence of a string being random if it falls within the given range. As exceeding the upper bound does not affect the randomness of the string, we ignore the upper bound and use the lower bound as a cutoff for a string being random, denoted by $\theta$, with a confidence by the proportion $\rho$: 
\begin{equation}
\theta = \mu\big(\hat{H}_{N}^{MLE}(w)\big) - t\times \sigma\big(\hat{H}_{N}^{MLE}(w)\big)
~~~~ ~~\rho = \frac{{number\ of\ samples\ above\ \theta}}{number\ of\ samples}
\end{equation}
Consequently, any strings falling below the threshold are considered not random, i.e. low-entropy blocks. Similarly, any strings falling above the threshold will be considered random, i.e. high-entropy. Table \ref{tbl:HNmu} shows thresholds ($\theta$) for $w$ using different window sizes (N) above a minimum level of confidence 99.0\%. 
\begin{table}[ht]
	\centering
    \begin{tabular}{|c|c|c|c|c|c|}
    \hline
    $~~~N~~~$    & $\mu$  & ~~$\sigma$ ~~&~~~~ $t$~~~~ & $\theta$ & $\rho$ \\ \hline
   16    &~~3.94199~~ &~~ 0.08290~~ &~~ 2.8~~  &~~ 3.7098~~ & ~~99.2\% ~~    \\ \hline
    32    & 4.88171 & 0.08134 & 2.7 & 4.6620 & 99.3\%     \\ \hline
    64    & 5.76562 & 0.07664 & 2.6  & 5.5663 & 99.2\%      \\ \hline
    128  & 6.55003 & 0.06733 & 2.5  & 6.3817 & 99.2\%      \\ \hline
    256  & 7.17518 & 0.05240 & 2.5  & 7.0441 & 99.2\%   \\ \hline
    512  & 7.59073 & 0.03364 & 2.4 & 7.5099 & 99.0\%            \\ \hline
    1024 & 7.80894 & 0.01726 & 2.5 & 7.7658 &99.1\%             \\ \hline
     2048 & 7.90804 & 0.00814 & 2.5 & 7.8877 & 99.2\%             \\ \hline
    \end{tabular}
    \caption {$\hat{H}_N^{MLE}(w)$ under Various Configurations}
	\label{tbl:HNmu}
\end{table}

\begin{figure}[ht]
	\begin{center}
	\includegraphics[scale=0.50]{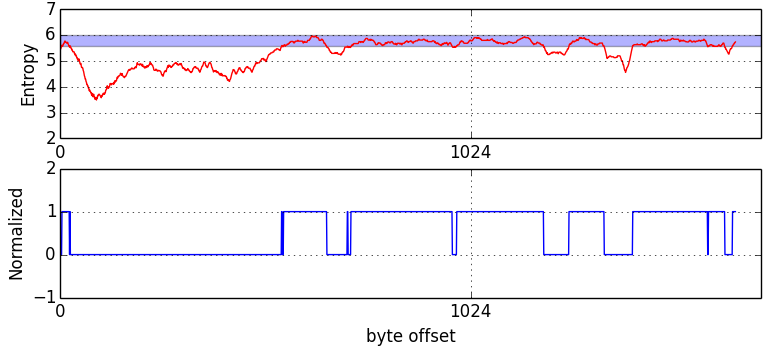}
	\caption{Normalization: high-entropy blocks}
	\label{fig:norm}
	\end{center}
\end{figure}

The confidence measures the confidence of a string not being random when falling out of the range, rather than a confidence of a string being random when falling within the range. For example, let $N$ be 64 and $\Sigma$=$\Sigma_0$, then $\mu$=5.7656, $\sigma$=0.0766. With 99.4\% of samples above $\theta$=$\mu$-3$\sigma$=5.53569 (i.e. $t = 3$), we would have \emph{at least} 99.4\% confidence that a string $s$ with $\hat{H}_N^{MLE}(s)$=5.5120 is not close to random, i.e. not a high-entropy block. Here, $t$ is our control variable. We can choose a smaller $t$ to tighten the range with a higher confidence or a larger $t$ to loosen the range, but with a lower confidence. In our study, we choose $t$ tightly to obtain a relatively high confidence, at least 99.0\%. %The following table shows various configurations with a confidence over 99.0\%.
With the threshold, we could then transform sample entropy score to either one or zero. The plot turns to be to a square wave where \emph{one} indicates high-entropy and \emph{zero} for low-entropy as shown in figure \ref{fig:norm}. The shadow in the upper plot shows the cutoff. 

\subsection{The choice of $\Sigma$}
\label{sigma}
\label{voting}

Due to statistical limitations, some data blocks may mistakenly be labeled as high-entropy blocks, i.e. a false positive, which will mislead the fingerprint and therefore must be avoided or minimized. In order to achieve this, we devised a voting mechanism using multi-resolution analysis, utilizing the choice of alphabet $\Sigma$. As will be shown, this mechanism dramatically reduces the rate of false positives.

Thus far we based our discussion on the choice of $\Sigma$ to be $\Sigma_0$ ($m$=256) with each char being a byte. In cryptography, however, the randomness of key material is defined at a more restrictive level, i.e. at a \textit{bit} level, and thereby $\Sigma$=\{0, 1\} ($m$=2). %Worth to note, our previous arguments will be not affected by such adjustment at all. 
Let's consider one experiment of tossing one coin that has two outcomes, and another experiment of tossing eight independent coins with two outcomes for each. According to basic probability theory, if each coin is uniformly drawn from $\Sigma$=\{0, 1\}, the outcome of eight coins ($\Sigma_0$) will still follow a uniform distribution. In our estimation of $\hat{H}^{MLE}_N(w)$, we do generate each random byte by randomly sampling eight times over \{0, 1\} for all our sample strings. That being said, given that each bit is independently sampled uniformly from \{0, 1\}, we could choose a random variable of different number of $\tau$ bits (i.e. coins) and such a random variable will be guaranteed to have a uniform distribution. 

As an extension to our previous computation of $\hat{H}^{MLE}_N(w)$, we outline the thresholds and their confidence levels for different $\tau$ while fixing $N$ to 32.  We use the term $\tau$-bit measure, e.g. 2-bit measure. Previously, $N$ could be interpreted as either the window size and the sample size. In the case of $\tau$-bit measure, the sample size changes, i.e. $\frac{8}{\tau}$ N ($\tau \leq 8$). For convenience, we abuse the notation N, using it as the window size in the rest of this paper. The use of $\tau$-bit measure does not change the fundamentals of \textit{N-truncated entropy} as it simply uses a larger sample size and a different alphabet. 
 
\begin{table}[ht]
	\centering
    \begin{tabular}{|c|c|c|c|c|c|c|}
    \hline
     $\tau$ & m   &  $\mu$     & $\sigma$ & $t$ & $\theta$ & $\rho$ \\ \hline
    ~~~1 ~~  &~~ 2~~ & ~~0.9971~~ &~~ 0.00399~~ &~~ 4.18 ~~ &~~ 0.9804 ~~& ~~99.28\%~~     \\ \hline
    2   & 4 & 1.9829 & 0.01387 & 3.59 & 1.9331 & 99.20\%     \\ \hline
    4   & 16 & 3.8196 & 0.06715 & 3.02  & 3.6168 & 99.31\%      \\ \hline
    8   & 256 & 4.8817 & 0.08135 & 3.0  & 4.6356  & 99.35\%      \\ \hline
    \end{tabular}
    \caption {$\tau$-bit measure $\hat{H}_{32}^{MLE}(w)$}
	\label{tbl:hmu2}
\end{table}

Statistical methods such as sample entropy generally ignore potential structures or patterns occurring in the data. Therefore, a string with a high sample entropy score is not guaranteed to be random. For example, given a hexadecimal string $s$ be ``55 55 bb bb'', i.e. $0101~ 0101~ 0101~ 0101~ 1010~ 1010~ 1010~ 1010$ in binary, we have $\hat{p}_0=\hat{p}_1=\frac{1}{2}$ if 1-bit measure ($\tau$=1) used, i.e. $\Sigma$=\{0, 1\}, and then $\hat{H}^{MLE}_N(s)$ = 1. Consequently, $s$ will be labeled as high-entropy bytes in spite they are not at all. Taking another example from real world, a hexadecimal string from a TLS session: $16~ 03~ 01~ 0c~ 13~ 0b~ 00~ 0c~ 0f~ 00~ 0d~ 0e~ 10~ 04~ 7a~ 30~ 82$, which is a block of control information\footnote{Control information is commonly known to have low entropy.} from the TLS handshake traffic. The two bytes \emph{03 01} indicate the TLS version, i.e. TLS 1.0, \emph{0c 13} for the length,\emph{0b} for the protocol type, and another 3 bytes of length \emph{00 0c 0f}. This block may not also appear ``random`` if an 8-bit measure is used. Such cases are prone to false positives and mislead the process.

\begin{figure}[ht]
	\begin{center}
	\includegraphics[scale=0.30]{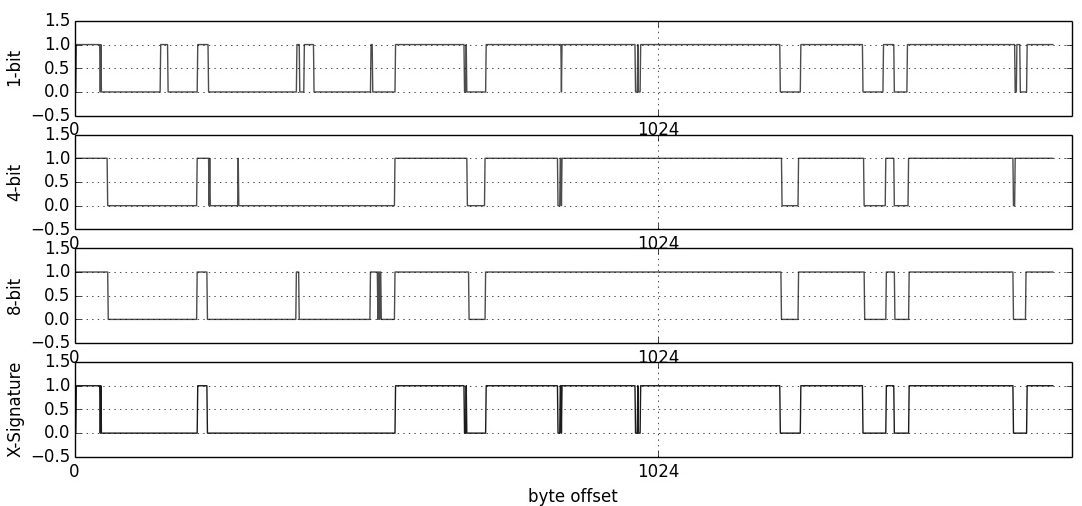}
	\caption{A traffic sample from a TLS 1.2 session with a 1024-bit RSA public key.}
	\label{fig:xsig}
	\end{center}
\end{figure}

However, the idea is that if a string is random, no matter which $\tau$-bit measure is being used, its sample entropy $\hat{H}^{MLE}_N(s)$ should be always close to $\hat{H}_N(\mathcal{U})$. Thus, we propose to use a voting mechanism instead of using a sole $\tau$-measure. The voting rule is if any of chosen $\tau$-bit measure rejects the randomness of that block, the block will be labeled as non-random. It is a simple AND operation among the outcome of all measures. Figure \ref{fig:xsig} shows the effectiveness of combining three $\tau$-measures, where the resulting signature by voting precisely outlines all high-entropy blocks in the TLS session. The last plot line, X-signature, is based on the voting over the three 1-bit, 4-bit and 8-bit measures.

\subsection{Filtering Threshold}

Our voting mechanism effectively reduces false positives. However, in some scenarios, this approach may still not be sufficient to eliminate all false positives. There is still a chance that all $\tau$-bit measures falsely identify an ordinary block to be high-entropy because of accidentally some small actual randomness within the data. If there supposedly are no high-entropy data blocks, the length of a data block with randomness should be less than the minimum length of interest and the size of detected high-entropy units would appear to be relatively small compared to that, if there actually exists a high-entropy data block of interest. A filtering threshold denoted as $\xi$ is possibly chosen to eliminate those small high-entropy units. Our empirical study suggests $\xi=9$ to be a good choice when a 32-byte sliding window size chosen for detecting a minimum 20-byte high-entropy key material blocks. That means if there are only 9 consecutive high-entropy blocks detected between two low-entropy units, then a false positive is identified and filtered out in that case. Here, the ``filter out'' means  labeling these blocks to be low-entropy instead of high-entropy. %Clearly, if the minimum length of interest is too small, this step will not be necessary. 

\subsection{Calibration}
\label{cali}
Beyond identifying high-entropy blocks, it is also essential to describe the length of each unit in order to fingerprint the shape of the square wave as shown in \ref{fig:xsig}. Due to its statistical inheritance and the way of measuring, the length of each unit (i.e. the number of detected consecutive high-entropy or low-entropy blocks) may vary because when the sliding window is partially over the target random bytes, it may still continue to yield high sample entropy blocks until the window moves sufficiently away from the target.  For example, a TLS traffic stream contains a client random number as a chunk of 28 bytes. It is not difficult to anticipate that there will not be only exactly one high-entropy block detected in this case. The total number of high-entropy blocks detected around that chunk of data will not be fixed as well from case to case. However, our intention is not to determine an absolute value for each unit among all cases, but rather a certain reasonable range. Hereby, we resort to \emph{Monte-Carlo} methods to empirically estimate the range. For example, to estimate the length of high-entropy unit around client random bytes, we sampled 100,000 \emph{client hello} messages from TLS sessions.

\begin{figure}[ht]
	\begin{center}
	\includegraphics[scale=0.15]{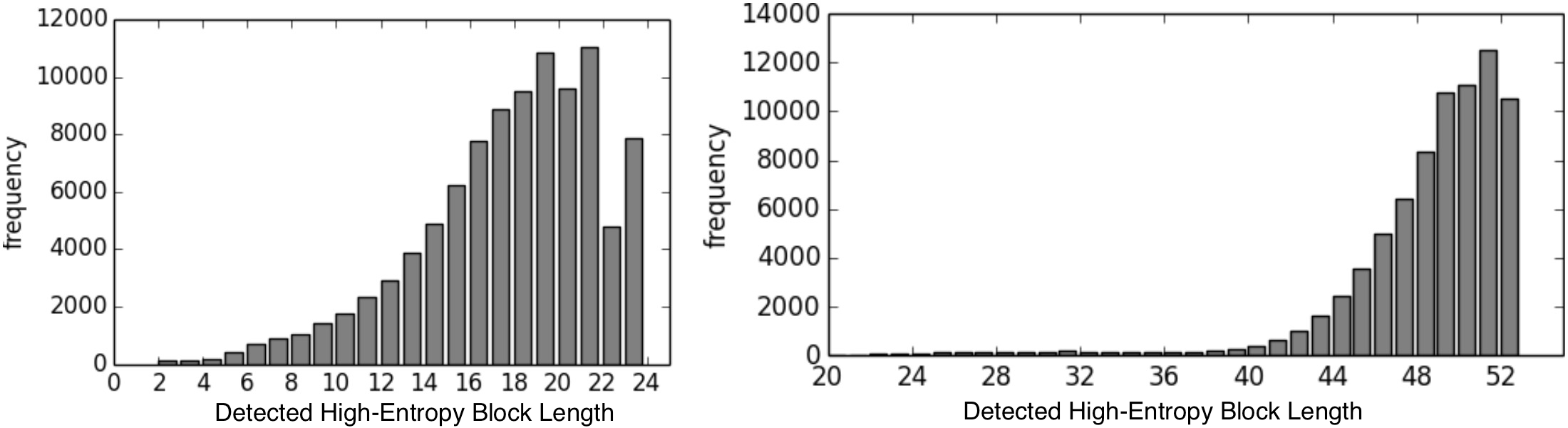}
	\caption{Distribution of length of detected high-entropy blocks (1) Left: over the TLS 28-byte client random string (2) Right: over the TLS 28-byte client random string and 32-byte session ID.}
	\label{fig:range}
	\end{center}
\end{figure}

The result shown in figure \ref{fig:range} indicates most of the length for the 28-byte client random string followed by the list of cipher suites fall within a range between six high-entropy blocks and twenty-four blocks. If a 32-byte TLS session ID (also random bytes) is present along with the client random bytes, adding up to 60 bytes, we obtain a range of $[38, 52]$ as shown in figure \ref{fig:range}. A more conservative range would be $[20, 52]$.

\subsection{Fingerprinting} % Key Exchange Protocol}

Fingerprinting is a process to profile a key exchange protocol by its distribution of high-entropy blocks along traffic streams generated by such a protocol. A entropy-based fingerprint is a series of interleaving high-entropy units and low-entropy units with the length of each unit specified as a range. The reason that high-entropy blocks have to interleave with low-entropy ones is that otherwise two adjacent high-entropy or low-entropy blocks would be merged into one. Let $(s, l, r)$ represent one unit where $s \in \{1, 0\}, l, r \in \mathbf{Z^+}$, where $s$ be the sign indicating a high-entropy unit or low-entropy, $l$ be the minimum length and $r$ be the maximum length. An entropy-based fingerprint then is the concatenation of an ordered list of $(s, l, r)$ with $s$ alternating among one and zero. Alternatively, it can be concisely expressed as below, where $s_i \in \{1, 0\}, l_i, r_i \in \mathbf{Z^+}$. The benefit of such a representation is that this form aligns with standard regular expression and the matching process can be done very efficiently. The regular expression form will provide a flexible way of expressing the fingerprint, for instance, optional units, as will be shown in the experiment section. 
\[
 \bigparallel_{i=1}^n s_i \{l_i, r_i\}, s_i \neq s_{i+1}
 \]
The fingerprinting is straightforward in three steps: (1) identify high-entropy and low-entropy areas (units) of the anticipated traffic from a cryptographic protocol; (2) follow the technique described in section \ref{cali} and estimate the range for each area; (3) formalize the units in a regular expression.  Taking TLS using a cipher-suite of DHE-RSA-* as an example, the fingerprint is as below:
\[
1\{8, 54\}0\{20,1024\}1\{8,54\}0\{30,800\}1\{80,260\} ....  
\]

During the detection phase, we have these steps: (1) scan the traffic stream by sliding a window over it and estimating sample entropy for each window using different $\tau$-bit measures; (2) normalize each block by its entropy score to either one or zero using the pre-calculated threshold $\theta$; (3) perform the voting (i.e. AND) of outcomes from each measures; (4) filter out the noises using filter threshold; (5) use regular expression to match the predefined fingerprint against the output (i.e. a string consisting of zeros and ones).
 
In our demonstration, we emphasize DHE-RSA-* cipher-suite for TLS protocol as our approach aims to profile that a particular key exchange protocol and TLS is capable of using different key exchange protocols. SSL has evolved over time into the standard TLS protocol, which supports a long list of cipher suites with different key exchange protocols. To demonstrate, we choose to profile one set of key exchange protocol cipher suites, i.e. DHE-RSA-*, (see \ref{tbl:cipher}). By contrast, as an application of our system, most botnet C\&C protocols are much simpler as most of them are designed for the sole purpose of performing a limited number of tasks.

\section{Evaluation}
\label{sec:eval}
SSL/TLS is a well-known cryptographic protocol with fair complexity. The successful characterization of the TLS protocol provides the full ability to characterize other and simpler botnet C\&C protocols. For evaluation, we first use TLS as our primary target and later extend it to the Nugache botnet. All streams are bidirectional and packets of a stream are correctly ordered with all TCP/IP headers removed. The \emph{tshark}~\cite{tshark} was used as a primary tool to process network traces in \texttt{pcap}~\cite{pcap}.

\subsection{Datasets}
%SF-JUL29
We obtained a data set of TLS network traffic from the ZMap project~\cite{zmap13}. Initially, we extracted 16,240 TCP streams on standard port 443 from 800MB of raw traffic data and further reduced to 5,794 completed and validated TLS streams\footnote{A large portion of hosts scanned by the ZMap client did not respond or reject connections for various reasons during TLS negotiation}. Then, we extracted from those 5,794 streams the 1,378 streams that used one of the DHE-RSA-$\ast$ ciphersuites in Table \ref{tbl:cipher}. We split 1,378 instances into two sets: the \emph{d00200} set of 218 instances for parameter selection and signature refinement and the test set \emph{d00300} of 1,160 instances for the testing of the final signature, denoted as the \emph{d00015} set. We also extracted 1,204 TLS instances with other ciphersuites. We extracted 337 Nugache traffic streams from a set of raw Nugache traffic and divided instances into two groups: 162 instances of training set and 175 instances of testing set. Similar to TLS, we use the training set to tune the fingerprint and the testing set for validation.

\begin{table}[ht]
	\centering
	\begin{tabular}{|c|l|}
	\hline 
	\textbf{Cipher} ID & \textbf{Name} \\ \hline \hline
   % 0x00004 & \verb|TLS_RSA_WITH_RC4128_MD5| \\ \hline
   % 0x00005  %\verb|TLS_RSA_WITH_RC4_128_SHA| \\ \hline
    0x00015 & \verb|TLS_DHE_RSA_WITH_DES_CBC_SHA| \\ \hline
    0x00016 & \verb|TLS_DHE_RSA_WITH_3DES_EDE_CBC_SHA| \\ \hline
    0x00033 & \verb|TLS_DHE_RSA_WITH_AES_128_CBC_SHA| \\ \hline
   % 0x00035 & \verb|TLS_RSA_WITH_AES_256_CBC_SHA| \\ \hline
    0x00039 & \verb|TLS_DHE_RSA_WITH_AES_256_CBC_SHA| \\ \hline
    0x00045 & \verb|TLS_DHE_RSA_WITH_CAMELLIA_128_CBC_SHA| \\ \hline
    0x00067 & \verb|TLS_DHE_RSA_WITH_AES_128_CBC_SHA256| \\ \hline
    0x0006B & \verb|TLS_DHE_RSA_WITH_AES_256_CBC_SHA256| \\ \hline
    0x00088 & \verb|TLS_DHE_RSA_WITH_CAMELLIA_256_CBC_SHA| \\ \hline
    0x0009A & \verb|TLS_DHE_RSA_WITH_SEED_CBC_SHA| \\ \hline
    0x0009E & \verb|TLS_DHE_RSA_WITH_AES_128_GCM_SHA256| \\ \hline
    0x0009F & \verb|TLS_DHE_RSA_WITH_AES_256_GCM_SHA384| \\ \hline 
	\end{tabular} 
	 \caption{TLS Ciphersuites of Choice: DHE-RSA-*}
	\label{tbl:cipher}
\end{table}
~\\
\begin{table}[ht]
	\centering
	\begin{tabular}{|c||c|c|c|c|c|c|c|c|c|}
	\hline
	\textbf{Port} & ~~80~~ &~~ 25~~ &~~ 22~~ & ~~143~~ &~~ 21~~ &~~ 111~~ & ~~179~~ &~~ 139~~ & ~~110~~  \\ \hline
    ~~\# of Streams ~~& 582 & 189 & 168 & 125 & 96 & 44 & 18 & 5 & 5 \\ \hline
    \end{tabular}
	%\caption{Part of Traffic Types by Port}
	\label{tbl:unswnb}
\end{table}
In addition, we used 3,412 non-TLS TCP streams from a data set generated by UNSW-NB15\cite{moustafa2015unsw}. This data set contains a variety of traffic types, but without any TLS traffic so we can use it as another dimension of negative cases for testing the fingerprints. Table above shows the traffic type of the majority by service ports, only including standard ports under 1024. The table does not show the whole spectrum of traffic types in this dataset, but rather provides a quick look. More details on this data set are available in the original paper. 

\subsection{TLS}

%\subsubsection{Initial Signature}

We test the signature generated as previously described over the training set \emph{dhe00200} with thresholds of the confidence $\rho$ above 99.2\% for different measures. The results shown in table \ref{tbl:origsig} do not seem promising at all, of all the best results from 1-4-8 and 1-2-4-8 only reach a recall rate, 62.84\% and 64.22\% with the confidence of 99.85\% respectively, but it does confirm that the strategy of using multiple $\tau$-measures significantly improves the recall rate. Also, it is interesting to notice that the rate of multiple $\tau$-measures drops significantly below 10\% with a confidence of 99.99\%, which is reasonable because the threshold is too relaxed (with a higher proportion of high entropy blocks) to be accurate. 

\begin{table}[ht]
	\centering
	\begin{tabular}{|c||c|c|c|c|c|c|c|}
	\hline
	\backslashbox{$\tau$}{$\rho$} & 99.20\% & 99.85\% & 99.97\% & 99.99\% \\ \hline\hline
	~~~1-bit ~~~& ~~8.72\% ~~& ~~36.70\% ~~& ~~26.15\% ~~& ~~26.14\% ~~\\ \hline
	2-bit & 15.13\% & 10.55\% & 23.39\% & 23.39\% \\ \hline
	4-bit & 47.25\% & 25.68\% & 8.26\% & 11.93\% \\ \hline
	8-bit & 42.40\% & 28.44\% & 3.21\% & 10.09\% \\ \hline
	1-2-8 & 31.19\% & 17.43\% & 7.80\% & 4.58\% \\ \hline
	1-4-8 & 45.41\% & \textbf{62.84\%} & 38.99\% & 5.50\% \\ \hline
	1-2-4-8 & 39.44\% & \textbf{64.22\%} & 39.44\% & 5.05\% \\ \hline
	    \end{tabular}
	   \caption{Recall rate of the original signature for TLS}
	\label{tbl:origsig}
\end{table}

%\subsubsection{Refinement}
By manually checking those failures, we found three major issues of our original signature. One is the range of the server random bytes. It was a little bit tighter than it appeared, which is previously set to be (+, 8, 54) as we used the range estimated from client random bytes. It turns out to be inadequate as the bytes after the server random bytes appear more random than those after client random bytes, and therefore more likely produce a longer high-entropy block. Following the same method as we did for client random bytes, we increase the maximum length to 64. The second major issue is that we failed to consider optional random bytes such as key identifier fields for both issuer and subject of the certificate. The third one relies on the fact that two high-entropy areas might be adjacent to each other without a sufficient gap and get merged to a larger high-entropy area, e.g. the signature of certificate and the server key exchange parameters. For the later two cases, we introduce optional blocks to the signature making the signature scalable. In the regular expression, we can include optional strings. For instance, our TLS signature has been extended to include optional strings as below. This adjustment boosts the recall for most cases, as shown in Table \ref{tbl:newsig}. For both cases of 1-4-8 and 1-2-4-8, the recall increases by around 20\%. 
{\small\[
 `...1\{80,260\}(0\{20,1024\}|\{8,160\}(1\{8,70\}|\{8,70\}0\{0,300\}1\{8,70\})0\{0,500\})...'. 
\]}

\begin{table}[ht]
	\centering
	\begin{tabular}{|c||c|c|c|c|c|c|c|}
	\hline
	\backslashbox{$\tau$}{$\rho$} & 99.20\% & 99.85\% & 99.97\% & 99.99\% \\ \hline\hline
	~~~1-bit~~~ & ~~12.39\%~~ &~~ 6.88\% ~~&~~ 40.37\% ~~&~~ 40.37\% ~~\\ \hline
	2-bit & 21.10\% & 19.27\% & 38.53\% & 40.83\% \\ \hline
	~~~4-bit ~~~& ~~84.40\% ~~&  ~~73.39\% ~~& 33.03\% ~~& 13.30\%~~ \\ \hline
	8-bit & 67.43\% & 51.83\% & 11.01\% & 16.51\% \\ \hline
	1-2-8 & 41.28\% & 25.69\% & 18.34\% & 11.93\% \\ \hline
	1-4-8 & 55.05\% & \textbf{82.57\%} & 67.43\% & 12.39\% \\ \hline
	1-2-4-8 & 49.54\% & \textbf{87.61\%} & 67.43\% & 12.39\% \\ \hline
	    \end{tabular}
	   \caption{Recall using refined fingerprint}
	\label{tbl:newsig}
\end{table}

\begin{figure}[ht]
	\begin{center}
	\includegraphics[scale=0.5]{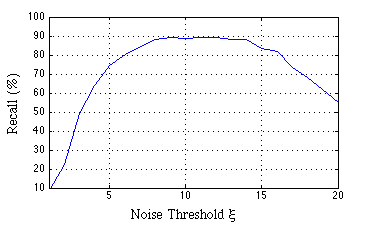}
	\caption{Noise Threshold Selection over TLS traffic using a 4-bit measure}
	\label{fig:4bit}
	\end{center}
\end{figure}

%\subsubsection{Filter Parameter $\xi$}
The noise threshold is used to remove false positives and make the fingerprint more reliable. As the threshold increases, the detection accuracy of high-entropy blocks will increase as we are eliminating those accidental ``high-entropy'' blocks. At a certain point, this elimination may hurt the effectiveness as true high-entropy blocks may be eliminated by such an excessively large threshold. We experimented with different filter thresholds $\xi$ using a 4-bit measure, as shown in figure \ref{fig:4bit}. Given its initial purpose, this parameter should be kept as small as possible for effective filtering. Thus, $\xi=9$ is chosen based on the empirical results. As suggested by our test results, it appears to be a proper choice for other measures, e.g. 1-4-8 measure.

\subsubsection{Test Results}
After two improvement procedures, i.e. signature refinement and parameter selection, the ultimate test over the testing sets, \emph{d00300}, is shown in the table below. The multiple $\tau$-measure 1-4-8 now produces a good recall rate. 
\begin{table}[ht]
    \centering
	\begin{tabular}{|c||c|c|c|}
	\hline
 	$\xi$=9	& ~~TP~~ &~~ FN~~ & ~~ Recall ~~\\ \hline\hline
	~~4-bit measure ($\rho$=99.20\%)~~ & 1056 & 104 & 91.03\% \\ \hline
	%4-bit (99.20\%, $\xi$=11) &1026 &134 & 88.45\%  \\ \hline
	1-4-8 measure ($\rho$=99.85\%) & 1079 & 81 & 93.02\%  \\ \hline
	%1-4-8 ($\xi$=10) &1076 & 84 & 92.75\%  \\ \hline
	%1-4-8 ($\xi$=11) &1073 & 87 & 92.50\%  \\ \hline
	   \end{tabular}
	\label{tbl:newsigtest}
\end{table}

%SF-JUL29

\begin{table}[ht]
	\centering
	\begin{tabular}{|l|c||c|c| c |}
	\hline
	~~~~~~ Dataset & ~~Total~~ & ~~Positive ~~& ~~Negative~~ \\ \hline
    ~~\emph{d00200}: TLS w/ selected Cipher~~ & 1,160 & 1,079 & 81 \\ \hline
    ~~\emph{d00300}: TLS w/ other Cipher & 1,204 & 61 & 1,143 \\ \hline
    ~~\emph{d00015}:  non-TLS  & 3,412 & 0 & 3,412 \\ \hline
    \end{tabular}
	\caption{TLS signature over different datasets }
	\label{tbl:signug}
\end{table}

Finally, we fixed our noise threshold $\xi=9$ and used the 1-4-8 measure. We summarize our results over three datasets as follows. Overall, the TLS signature has a precision of nearly 94.6\% and its accuracy is around 94\%, only including negative cases from \emph{d00300} so as to have a equivalent size of positive cases. On the other hand, negative cases from non-TLS, i.e. \emph{d00015}, turn out to be relatively trivial even though some of instances do contain high-entropy traffic, for example, SSH on port 22. 

%As the result suggests in Table , it is preferable to choose a smaller noise threshold when it can achieve an relatively equivalent recall. A proper noise threshold may effectively eliminate stubborn noise. However, a loose noise threshold may also mistakenly slice out those small actual high-entropy blocks, which may cause false positives. In the case of TLS, $\xi$=9 seems to be a good choice.

\subsection{Application on Botnet Detection: Nugache}

The Nugache botnet, was one of the first peer-to-peer botnets to use strong cryptography to protect its C\&C channel, as the inter-peer communication was encrypted using individually negotiated session keys derived using a hybrid RSA/Rijndael scheme \cite{sok:p2pwned,dd2008:malware,art:stormnugache}. Specifically, Nugache uses a two-way RSA-like key exchange protocol for every session with a minimum length of 512 bits for the modulus. That is, one peer sends the length of the key to announce a peer key exchange, followed by an actual key~\cite{dd2008:malware}; the other peer in turn replies with a message of the same length encrypted with that public key. Compared to TLS, signature extraction for Nugache is much easier because of the simplicity of its key exchange. Since there is little control information in key exchange messages, if consider the payload only, the signature can be simply defined as 1*, meaning high-entropy blocks everywhere, which is also a strong detectable characteristic distinct from other cryptographic protocols. Following the same consideration, we choose $\xi=9$, which yields a fair recall rate.
%\subsubsection{Filter Parameter $\xi$} 
\begin{table}[ht]
	\centering
	\begin{tabular}{|c||c|c|c|c|c|c|c|}
	\hline
	\backslashbox{$\tau$}{$\rho$} & 99.20\% & 99.85\% & 99.97\% & 99.99\% \\ \hline\hline
%	~~~1-bit ~~~&~~ 85.80\% ~~&~~ 74.69\% ~~&~~ 40.12\% ~~&~~ 40.12\%~~ \\ \hline
	2-bit & \textbf{92.21\%} & 67.90\% & 39.50\% & 17.28\% \\ \hline
%	4-bit & 64.81\% & 30.86\% & 13.58\% & 0.0\% \\ \hline
%	8-bit & 69.14\% & 61.73\% & 33.95\% & 11.72\% \\ \hline
	1-2-8 & 88.27\% & \textbf{90.12\%} & 73.46\% & 56.17\% \\ \hline
	1-4-8 & 89.51\% & \textbf{92.21\%} & 75.93\% & 56.17\% \\ \hline
	1-2-4-8 & \textbf{90.12\%} & \textbf{95.06\%} & 77.16\% & 56.17\% \\ \hline
	    \end{tabular}
	   \caption{Recall on Nugache (N=32)}
	\label{tbl:signug}
\end{table}

 %The choice of $\xi$ coincides with the one used for TLS. 
%\begin{figure}[h!]
%	\begin{center}
%	\includegraphics[scale=0.5]{./b2xi.png}
%	\caption{Noise Threshold Selection over Nugache using a 2-bit measure}
%	\label{fig:2bit}
%	\end{center}
%\end{figure}
%\subsubsection{Testing}
The initial fingerprint we generated for Nugache includes two high-entropy areas, corresponding to the two-way key exchange. First, we test all $\tau$-bit measures with a fixed noise threshold value $\xi$=9. It shows the 2-bit measure produces good results (92.21\% with $\rho=99.20\%$) but meanwhile our voting mechanism clearly outperforms a single $\tau$-bit measure given the same level of confidence. We conservatively choose the 1-4-8 measure as our metric in a general. 

% \begin{table}[ht]
% 	\scriptsize
% 	\centering
% 	\begin{tabular}{|c||c|c|c|c|c|c|}
% 	\hline
% 	\backslashbox{$\tau$}{$\rho$} & 99.20\% & 99.85\% & 99.99\% \\ \hline\hline
% 	1-bit & 40.65\% & 12.76\% & 2.96\% \\ \hline
% 	2-bit & 45.69\%& 19.29\% & 2.96\% \\ \hline
% 	4-bit & 44.80\% & 17.50\%  & 0.0\% \\ \hline
% 	8-bit & 66.17\% & 33.83\% & 2.67\% \\ \hline
% 	1-2-8 & \textbf{94.95\%} & 81.31\%  & 37.98\% \\ \hline
% 	1-4-8 & \textbf{94.95\%} & 83.38\% & 39.17\% \\ \hline
% 	1-2-4-8 & \textbf{94.95\%} & 84.27\% & 40.95\% \\ \hline
% 	    \end{tabular}
% 	   \caption{\scriptsize Recall of all measures on Nugache (N=64) }
% 	\label{tbl:signug}
% \end{table}

%SF-JUL29
\begin{table}[ht]
	\centering
	\begin{tabular}{|c|c||c|c| c |}
	\hline
	Dataset & Desc & ~~Total~~ & ~~Positive~~ & ~~Negative~~ \\ \hline
	  - & ~~\emph{Nugache}~~ & 175 & 162 & 13 \\ \hline
   ~~ \emph{d00200, d00300} ~~& TLS & 2,364 & 0 & 2,364 \\ \hline
    \emph{d00015} &  non-TLS  & 3,412 & 0 & 3,412 \\ \hline
    \end{tabular}
	\caption{Nugache fingerprint over different datasets }
	\label{tbl:finalug}
\end{table}
%We use the same results from our previous experiments with the fixed noise threshold $\xi=9$ and the 1-4-8 measure. 
In Table \ref{tbl:finalug}, we summarize our testing results of the Nugache signature over three datasets as follows. It is encouraging that the Nugache signature generates no false positive and so has a precision of 100\%. For obfuscation techniques, there still a portion of the traffic, although small, will appear to have low entropy. 

\section{Limitations \& Future Work}

%As signature-based, this approach shares similar shortcomings of other signature-based methods. However, if the signature generation process is automated, and new signatures can be easily generated and deployed so long as sufficient traffic instances available, then it would be still valuable for such approaches.

One may argue that high entropy does not necessarily imply encryption, compressed data, or multimedia data. The critical point is the distribution of high-entropy data blocks not solely the presence of high-entropy data. A study \cite{ZhangPM13} provides evidence against such ``common sense,'' where it was shown that multimedia files could yield low entropy instead, although the authors also pointed out that in some cases compressed files do have high entropy. Such cases require a much closer look, which we left for future work. Furthermore, encodings, e.g. base64~\cite{base64:rfc4648}, can significantly reduce the entropy of a string. For this case, we assume that a base64 detector as well as a decoder could be deployed to canonicalize the traffic data. It is also possible that one could easily inject arbitrary bytes to disturb the original distribution of high entropy and low entropy. In this case, we consider it to be a new protocol for which the traffic could be possibly fingerprinted, e.g. using optional units as we did for TLS. If the signature generation process is automated, then this approach would still be efficient. However, if more advanced obfuscation techniques~\cite{obfsproxy, ccs2013-fte} are applied, then our approach will fail at identifying the obfuscated protocol. Nevertheless, our proposed techniques may be still used to detect the obfuscation techniques themselves. 
% Instead of a customized cryptographic protocol for key exchange, an alternative for malware or botnets would be to adopt TLS as a cryptographic protocol. As discovered in a malware study\cite{anderson2016deciphering}, 10\% of malware samples indeed utilize TLS. Moreover, that

To avoid being fingerprinted, malware could adopt plain TLS instead of customizing the protocol, running the risk of SSL inspection. It may explain why there only 10\% of malware samples indeed utilize TLS. Nevertheless, the work \cite{anderson2016deciphering} also found that malware or botnets utilize TLS in a very customized way, i.e. advertising significantly much fewer cipher suites than enterprise TLS clients. A shorter list of cipher suites will reduce the control information (i.e. low-entropy blocks) and therefore may end with different fingerprints than enterprise-grade TLS clients. Investigating how effective our approach would be in such a scenario is left for future work. Under certain circumstances, it is possible that our approach may not be sufficient to rule out all possible false positives and we would recommend to coordinate with other tools for reducing false positives.

Last but not least, we are interested in looking at more diverse data, such as compressed data, SSH, and other malware traffic. 

%Another way botnet master may use to defeat our system is to use data expansion function, like pseudorandom function (PRF). PRF accepts a short random string and outputs a longer pseudorandom string. By using PRF, botnet master might enable bots to exchange a short key materials, 

%From \cite{Dorfinger2011TMA}, \emph{``Two concepts within data processing result in a high value of entropy. First data compression, as the bits needed for data representation should be minimized. Second data encryption, as any predictable behaviour available in the source data has to be removed. Both processing steps end with a data stream with equal probabilities for each event/symbol.``}

%Traffic Morphing
% FTE \cite{ccs2013-fte} SkypeMorph \cite{skypemorph12}
%(This may be a hard one to defeat)

%\section{Future Work}
%
%automate\\
%signature with low entropy\\
%more scalabile\\
%more botnet instances to evaluate

\section{Conclusion}
In this paper, we proposed a novel voting-based method for accurately detecting high-entropy blocks, e.g. key material, in a network traffic stream, and a method based on regular expressions for generating a scalable fingerprint based on identified high-entropy blocks. Our approach can effectively put malware authors on the defense, as a longer key used for a more securely encrypted connection would make it more easily characterized and therefore more detectable. However, if a shorter key is used for making the connection less vulnerable to detection, then they would only achieve a less secure connection.

%
% ---- Bibliography ----
%
%\begin{thebibliography}
\bibliographystyle{abbrv}
\bibliography{entropy_arxiv.bbl}

\begin{thebibliography}{10}

\bibitem{anderson2016deciphering}
B.~Anderson, S.~Paul, and D.~McGrew.
\newblock Deciphering malware's use of {TLS} (without decryption).
\newblock {\em arXiv preprint arXiv:1607.01639}, 2016.

\bibitem{Antos2001}
A.~Antos and I.~Kontoyiannis.
\newblock Convergence properties of functional estimates for discrete
  distributions.
\newblock {\em Random Struct. Algorithms}, 19(3-4):163--193, Oct. 2001.

\bibitem{tshark}
G.~Combs.
\newblock Wireshark: The network protocol analyzer.
\newblock http://www.wireshark.org/, 2016.

\bibitem{tls:rfc5246}
T.~Dierks and E.~Rescorla.
\newblock {RFC 5246: The Transport Layer Security (TLS) Protocol Version 1.2}.
\newblock http://tools.ietf.org/html/rfc5246, August 2008.

\bibitem{obfsproxy}
R.~Dingledine.
\newblock Obfsproxy: The next step in the censorship arms race.
\newblock {\em https://blog.torproject.org/blog/
  obfsproxy-next-step-censorship-arms-race}, 2012.

\bibitem{dd2008:malware}
D.~Dittrich and S.~Dietrich.
\newblock {P2P} as botnet command and control: a deeper insight.
\newblock In {\em Proceedings of the 3rd International Conference on Malicious
  and Unwanted Software (Malware)}, 2008.

\bibitem{Dorfinger2011TMA}
P.~Dorfinger, G.~Panholzer, and W.~John.
\newblock Entropy estimation for real-time encrypted traffic identification.
\newblock In {\em Proceedings of the Third International Conference on Traffic
  Monitoring and Analysis}, TMA'11, pages 164--171, Berlin, Heidelberg, 2011.
  Springer-Verlag.

\bibitem{zmap13}
Z.~Durumeric, E.~Wustrow, and J.~A. Halderman.
\newblock {ZMap}: {F}ast {I}nternet-wide scanning and its security
  applications.
\newblock In {\em Proceedings of the 22nd {USENIX} Security Symposium}, August
  2013.

\bibitem{dyer2013protocol}
K.~P. Dyer, S.~E. Coull, T.~Ristenpart, and T.~Shrimpton.
\newblock Protocol misidentification made easy with format-transforming
  encryption.
\newblock In {\em Proceedings of the 20th ACM SIGSAC conference on Computer and
  Communications Security}, pages 61--72. ACM, 2013.

\bibitem{ccs2013-fte}
K.~P. Dyer, S.~E. Coull, T.~Ristenpart, and T.~Shrimpton.
\newblock Protocol misidentification made easy with format-transforming
  encryption.
\newblock In {\em Proceedings of the 20th ACM conference on Computer and
  Communications Security (CCS 2013)}, November 2013.

\bibitem{jackknife}
B.~Efron and C.~Stein.
\newblock The jackknife estimate of variance.
\newblock {\em The Annals of Statistics 9 (1981)}, 9:586--596, April 2007.

\bibitem{ssh:rfc4251}
A.~Freier, P.~Karlton, and P.~Kocher.
\newblock {RFC 4251: The Secure Shell (SSH) Protocol Architecture}.
\newblock http://http://tools.ietf.org/html/rfc4251, Jan 2006.

\bibitem{ssl:rfc6101}
A.~Freier, P.~Karlton, and P.~Kocher.
\newblock {RFC 6101 (historic): The Secure Sockets Layer (SSL) Protocol Version
  3.0}.
\newblock http://tools.ietf.org/html/rfc6101, August 2011.

\bibitem{bothunter2007usenix}
G.~Gu, P.~Porras, V.~Yegneswaran, M.~Fong, and W.~Lee.
\newblock {BotHunter: Detecting Malware Infection Through IDS-driven Dialog
  Correlation}.
\newblock In {\em Proceedings of the 16th USENIX Security Symposium}, pages
  1--16, Berkeley, CA, USA, 2007. USENIX Association.

\bibitem{pcap}
V.~Jacobson, C.~Leres, and S.~McCanne.
\newblock tcpdump, a powerful command-line network analyzer.
\newblock http://www.tcpdump.org.

\bibitem{base64:rfc4648}
S.~Josefsson.
\newblock {RFC4648: The Base16, Base32, and Base64 Data Encodings}.
\newblock http://tools.ietf.org/html/rfc4648, 2006.

\bibitem{Kara:2005}
T.~Karagiannis, K.~Papagiannaki, and M.~Faloutsos.
\newblock Blinc: Multilevel traffic classification in the dark.
\newblock In {\em Proceedings of the 2005 Conference on Applications,
  Technologies, Architectures, and Protocols for Computer Communications},
  SIGCOMM '05, pages 229--240, New York, NY, USA, 2005. ACM.

\bibitem{miller}
G.~A. Miller.
\newblock Note on the bias of information estimates.
\newblock {\em Information Theory in Psychology: Problems and Methods}, pages
  pp. 95--100, 1955.

\bibitem{moustafa2015unsw}
N.~Moustafa and J.~Slay.
\newblock {UNSW-NB15: a comprehensive data set for network intrusion detection
  systems (UNSW-NB15 network data set)}.
\newblock In {\em Military Communications and Information Systems Conference
  (MilCIS), 2015}, pages 1--6. IEEE, 2015.

\bibitem{netentropy}
J.~Olivain and J.~Goubault{-}Larrecq.
\newblock Detecting subverted cryptographic protocols by entropy checking.
\newblock Research Report LSV-06-13, Laboratoire Sp{\'e}cification et
  V{\'e}rification, ENS Cachan, France, June 2006.
\newblock 19~pages.

\bibitem{Paninski03}
L.~Paninski.
\newblock Estimation of entropy and mutual information.
\newblock {\em Neural Comput.}, 15(6):1191--1253, June 2003.

\bibitem{dhe:rfc2631}
E.~Rescorla.
\newblock {RFC2631: Diffie-Hellman Key Agreement Method}.
\newblock https://tools.ietf.org/html/rfc2631, 1999.

\bibitem{sok:p2pwned}
C.~Rossow, D.~Andriesse, T.~Werner, B.~Stone-Gross, D.~Plohmann, C.~Dietrich,
  and H.~Bos.
\newblock {SoK}: {P2PWNED} - modeling and evaluating the resilience of
  peer-to-peer botnets.
\newblock In {\em 2013 IEEE Symposium on Security and Privacy}, pages 97--111,
  May 2013.

\bibitem{Schuermann04biasanalysis}
T.~Sch\"{u}rmann.
\newblock Bias analysis in entropy estimation.
\newblock {\em J. Phys. A Math. Gen}, pages 295--301, 2004.

\bibitem{shannon1948}
C.~E. Shannon.
\newblock A mathematical theory of communication.
\newblock {\em The Bell System Technical Journal}, 27:379--423, 623--656, July,
  October 1948.

\bibitem{art:stormnugache}
S.~Stover, D.~Dittrich, J.~Hernandez, and S.~Dietrich.
\newblock {Analysis of the Storm and Nugache Trojans: P2P is here}.
\newblock In {\em USENIX ;login: vol. 32, no. 6}, December 2007.

\bibitem{wang2015seeing}
L.~Wang, K.~P. Dyer, A.~Akella, T.~Ristenpart, and T.~Shrimpton.
\newblock Seeing through network-protocol obfuscation.
\newblock In {\em Proceedings of the 22nd ACM SIGSAC Conference on Computer and
  Communications Security}, pages 57--69. ACM, 2015.

\bibitem{white2013clear}
A.~M. White, S.~Krishnan, M.~Bailey, F.~Monrose, and P.~A. Porras.
\newblock Clear and present data: Opaque traffic and its security implications
  for the future.
\newblock In {\em NDSS}, 2013.

\bibitem{Wright:2006}
C.~V. Wright, F.~Monrose, and G.~M. Masson.
\newblock On inferring application protocol behaviors in encrypted network
  traffic.
\newblock {\em J. Mach. Learn. Res.}, 7:2745--2769, Dec. 2006.

\bibitem{tamd}
T.-F. Yen and M.~K. Reiter.
\newblock Traffic aggregation for malware detection.
\newblock In {\em Detection of Intrusions and Malware, and Vulnerability
  Assessment (DIMVA)}, pages 207--227. Springer, 2008.

\bibitem{zhang15}
H.~Zhang and C.~Papadopoulos.
\newblock Early detection of high entropy traffic.
\newblock In {\em {IEEE Conference on Communications and Network Security
  (CNS)}}, pages 104--112, Sept 2015.

\bibitem{ZhangPM13}
H.~Zhang, C.~Papadopoulos, and D.~Massey.
\newblock Detecting encrypted botnet traffic.
\newblock In {\em Proceedings of the IEEE INFOCOM 2013, Turin, Italy, April
  14-19, 2013}, pages 3453--1358, 2013.

\end{thebibliography}

\end{document}